\documentclass[onecolumn,amsmath,amssymb,aps,pra]{revtex4-2}

\usepackage{graphicx}
\usepackage{dcolumn}
\usepackage{bm}
\usepackage{physics}
\usepackage{hyperref}

\begin{document}
	
	\title{Travelling Dark State Polariton as a Viable Quantum Memory in a Solid-State Medium}
	
	\author{Avirup Chakraborty}
	\affiliation{S.N. Bose National Centre for Basic Sciences, \\JD Block, Sector-III, Bidhannagar, Kolkata, West Bengal-700106, India.}
	
	\author{Shrabana Chakrabarti}
	\affiliation{Sister Nivedita University, \\DG 1/2, DG Block(Newtown), Action Area I, Newtown, Kolkata, Chakpachuria, West Bengal 700156, India.}
	
	\date{\today}
	
	\begin{abstract}
		We theorize a quantum memory based on the dark-state polariton field, formed by the superposition of atomic and photonic states of a travelling probe laser pulse under the application of standing wave modes of a dominant control laser pulse using a lambda-level scheme Electromagnetically Induced Transparency in a solid medium. We show how an enhancement in the storage time for the pulse is achieved by eliminating pulse broadening due to diffusion. At last, we propose an experiment that can help realize the storage of a probe pulse in the hyperfine levels $^3\text{H}_4 \leftrightarrow ^1\text{D}_2$ of $\text{Pr}^{3+}:\text{Y}_2\text{SiO}_5$, cryogenically cooled at $4.5\text{ K}$. We also discuss multiple applications the storage of the quantum states the probe pulse with a prolonged time interval must have.
	\end{abstract}
	
	\keywords{Electromagnetically Induced Transparency, Quantum Memory, Dark State Polariton, Quantum Computation}
	
	\maketitle
	
	\section{Introduction}
	
	The promise shown by quantum information as a means for computation is immense. This is primarily because of the characteristic of a quantum system. The size and complexity of the wave function, a mathematical function which represents a quantum state, can grow exponentially large as the number of particles in the quantum system increases. This exponential size can help us represent complex problems that a classical computer cannot \cite{krylov2021}. Feynman was perhaps the first person to realize how much significance information processing in a quantum computer might hold in the future. He theorized how quantum laws can be simulated and understood at a deeper level in a quantum computer \cite{feynman1982}. Any quantum algorithms of the future would require reliable, robust, and long-term storage solutions of quantum states to represent information in order to process and have any significant impact on the computational landscape. It is, however, very difficult in practice to maintain a stable quantum state over a long period of time due to quantum decoherence, pertaining to the environmental interference with a quantum system. It has been found that quantum states of photons of light are more reliable for storage purposes than any other quantum states, because of their relatively better isolation from the environment and ease of manipulation for storage of information \cite{obrien2003, petrosyan2005}. In this paper, we put forward a novel approach of storing and retrieving quantum information by using a quasi-particle polariton state, a mixed atomic dipole and photonic state generated by strong atom-light interaction in an electromagnetically induced transparent (EIT) ultra-cold atomic lattice ensemble.
	
	Initial attempts at storing photonic states were very challenging and limited, particularly because of the fact that any quantum state suffers from spontaneous decay from its initial state. The decoherence in turn makes it challenging to use any such quantum state to encode information. One of the first attempts to use quantum states to store information was done by storing states of atoms in a Zeeman ground state by coupling atomic levels with a cavity QED field and subsequent adiabatic transfer of the information of the atomic states to the states of cavity QED photons \cite{parkins1993}. However, the cavity damping of the QED field remained a practical issue preventing the realization of a robust quantum memory architecture. The cavity damping and population loss from the atomic levels storing a quantum state was shown to be bypassed by using a three-level atomic system and two coherent lasers coupling an intermediate state with lowest lifetime to an initial atomic state (in which the system is initially prepared) and a final atomic state (to which the system evolves at the end of the process). Both states have a much longer lifetime than the intermediate state. This process leads the populations from the initial state to the intermediate state, which then get transferred to the final state by a Stokes pulse, bypassing the spontaneous emission from the intermediate state \cite{bergmann1998}. 
	
	The three-level scheme to store quantum information as quantum states of an atom in an energy state was further extended to the collective atomic ensemble by the phenomenon of electromagnetically induced transparency (EIT). In this process, a quasi-quantum excitation field, composed of Raman-like atomic spin coherence states and photonic states, was generated inside an atomic medium by coupling a weaker ``probe'' and a stronger ``control'' laser between an initial ground state (at which the system is maintained initially) and a final stable ground state (to which the system evolves eventually) to an excited state for coherent Raman excitation of the populations. The coupling between the stable state and the excited state facilitated by the control laser splits the excited state, due to which the probe laser does not get absorbed by the atomic medium; instead, the photonic states of the laser get transferred to the collective atomic excitation of the medium, culminating in the stable state creating a `dark state'. The dark state is a superposition state of the initial state and the final stable state and therefore remains protected from spontaneous emission. When such an atomic ensemble couples to a strong control cavity QED field and reaches the collective dark state, the quantum state of the probe pulse gets stored in the atomic spin excitation of the final stable ground state, creating a superposition of the photonic and the atomic spin excitation state, known as `dark state polariton' \cite{fleischhauer2000}. 
	
	The coherence of the spin atomic excitation was shown to be enhanced by manipulation of the dispersion of the atomic medium via the application of two counter-directional control lasers coupling the final stable state with the excited state. By the effect of these lasers, a standing wave is generated in the atomic medium, which creates a bandgap in the band structure of the atomic medium, facilitating trapping of the probe pulse inside the medium as a standing wave probe \cite{andre2002}. The effect of the standing wave probe inside the medium showed a greater coherence time for the quantum state stored inside it and hence provided a window of possibility to have more stable and robust quantum memories \cite{bajcsy2003, zimmer2006, andre2005}. It was theoretically shown how the counter-propagating control lasers eliminate total dispersion loss of the quantum state stored as a perfectly standing wave probe field inside an EIT medium \cite{hansen2007}.
	
	The trapped standing wave probe field can be a potent quantum memory in future quantum computers. The retrieval of information as a photonic state of the probe, however, requires one of the counter-directional lasers to be shut off for the probe to be retrieved by a photon detector in the desired direction. It can be inferred that the elimination of the dispersion loss will not be present while the retrieval of the stored standing wave pulse as a travelling wave pulse is in progress inside the atomic medium. So, in fact, the retrieved information about the quantum state will still not be free from environmental interference and will suffer incoherent losses. Therefore, the retrieval and propagation of the probe pulse inside the atomic medium as a polariton excitation also demands shielding from the eminent diffusion losses due to the diffusive motion of the atoms. It has already been shown that counter-directional control laser driven EIT systems in the frequency-up conversion regime provide much better transparency windows to travelling probe pulses inside the medium than a travelling wave control laser field. 
	
	In this paper, we put forward an adiabatic technique to address the issue by extending the shielding effect of the standing wave control field from the dispersion losses in the standing wave probe inside an EIT medium to the travelling probe as well, and we compare their effects with the travelling wave control field case. We present an adiabatic Raman population transfer technique of EIT by incorporating the effects of a standing wave control laser QED field on a travelling wave probe in an ensemble composed of stationary atoms. We derive the propagation equation for the travelling probe field inside the EIT medium and show how the Fourier modes of the standing wave control laser field eliminate the diffusion loss in the travelling probe as well. We also propose an experiment to show how the storage of a travelling probe pulse can be realized in a stationary atomic medium such as a rare-earth doped crystal lattice of $\text{Pr}^{3+}:\text{Y}_2\text{SiO}_5$ by creation of an EIT system, using a Dye laser to couple the hyperfine levels $^3\text{H}_4 \leftrightarrow ^1\text{D}_2$ of the crystal, having a very long lifetime.
	
	\section{Methodology}
	
	Our system is implemented using a solid crystalline atomic medium, structured as an ensemble lattice of atoms. We employ a lambda-level Electromagnetically Induced Transparency (EIT) scheme, wherein two hyperfine components of the ground-state energy levels are connected to an excited atomic state. Specifically, two strong counter-propagating coupling lasers, referred to as control lasers, with Rabi frequency $\Omega_c$, couple the final stable ground-state component $|c\rangle$ to the excited state $|a\rangle$. Concurrently, a weak coupling laser, referred to as the probe laser, links the initial ground-state component $|b\rangle$ with the same excited state $|a\rangle$. Notably, the direct transition between the two ground-state components $|b\rangle$ and $|c\rangle$ is forbidden. 
	
	The system leverages quantum interference effects: the destructive interference of transition probability amplitudes for the transitions $|b\rangle \to |a\rangle$ and $|c\rangle \to |a\rangle$. This interference enables the probe laser pulse to propagate through the medium without significant absorption. Furthermore, the wave vectors of the probe and control lasers are configured to satisfy the velocity-tuned multiphoton processes, ensuring coherent coupling and efficient EIT dynamics. This configuration is crucial for achieving the desired transparency and enhanced control over light-matter interactions within the medium.
	
	\begin{figure}[htbp]
		\centering
		\includegraphics[width=0.5\linewidth]{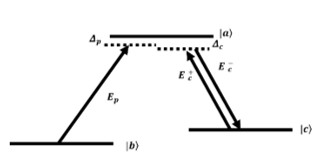}
		\caption{A Lambda Level EIT System, where counter-propagating control lasers with amplitude $E_c^\pm$ and a probe laser with amplitude $E_p$; $\Delta_p$ and $\Delta_c$ are the detunings.}
		\label{fig:fig1}
	\end{figure}
	
	\begin{align}
		\Delta_p - \mathbf{v} \cdot \mathbf{k}_p &= q \mathbf{k}_c \cdot \mathbf{v}, \quad q=0, \pm 2, \dots \nonumber \\
		\Delta_p - \mathbf{v} \cdot \mathbf{k}_p &= -\Delta_c + q \mathbf{k}_c \cdot \mathbf{v}, \quad q=0, \pm 1, \dots
	\end{align}
	
	Here, $q$ represents the number of photons in the control laser pulse, while $\mathbf{v}$ denotes the atomic velocity. The detunings of the probe and control laser pulses are denoted by $\Delta_p$ and $\Delta_c$, respectively. We consider a system comprising $N$ lambda-level atoms interacting with two distinct laser fields: a standing-wave control laser field with amplitude $E_c^\pm$ and wave vector $\pm \mathbf{k}_c$, and a traveling-wave probe field with amplitude $E_p$ and wave vector $\mathbf{k}_p$. The atomic states are described using density matrix elements $\sigma_{\mu\nu}$, while $d_{\mu\nu}$ denotes the dipole operators. 
	
	For simplicity, we assume the atomic velocity $\mathbf{v}$ to be approximately zero, consistent with the ensemble configuration of our atomic lattice system. However, we acknowledge the presence of atomic diffusion in real lattice systems, where $\mathbf{v}$ cannot be strictly zero. To further simplify the model, we consider near-resonant conditions for both the probe and control laser fields, such that $\Delta_p \approx 0$ and $\Delta_c \approx 0$. Under these assumptions, the Hamiltonian describing the interaction of the $N$-atom lambda-level system is expressed as follows:
	\begin{align}
		H &= \hbar\omega_a\sigma_{aa} + \hbar\omega_b\sigma_{bb} + \hbar\omega_c\sigma_{cc} - E_p e^{-i(\mathbf{k}_p \cdot \mathbf{r} - \omega_p t)} d_{ba}\sigma_{ba} \nonumber \\
		&\quad + \left(E_c^+ e^{+i(\mathbf{k}_c \cdot \mathbf{r} - \omega_c t)} + E_c^- e^{-i(\mathbf{k}_c \cdot \mathbf{r} - \omega_c t)}\right) d_{ca}\sigma_{ca} + \text{H.c.} \tag{1}
	\end{align}
	
	This formulation provides a foundation for analysing the coherent dynamics and light-matter interactions within the atomic lattice under the specified EIT scheme. In the presence of laser-atom interaction, we assume that the time evolution of the atomic operators is slow compared to the timescale of the electromagnetic fields. This allows us to define slowly varying atomic elements for the analysis. Specifically, we define $g_c = (d_{23} \cdot \mathbf{e}_c)/\hbar$ and $g_p = (d_{13} \cdot \mathbf{e}_p)/\hbar$, where $\mathbf{e}_c$ and $\mathbf{e}_p$ are unit vectors along the propagation directions of the control and probe lasers, respectively. Additionally, the slowly varying atomic density matrix elements are denoted as $\sigma_{\mu\nu}$. Substituting the slowly varying atomic elements into Heisenberg's equation of motion and applying the rotating wave approximation (RWA), which eliminates rapidly varying terms, we derive the Heisenberg-Langevin equations of motion. Here, $\Upsilon$ and $\Gamma_{\mu\nu}$ represent the incoherent decay rates of the electron population in specific states.
	\begin{align}
		\dot{\sigma}_{aa} &= -ig_p E_p e^{i\mathbf{k}_p \cdot \mathbf{r}} \sigma_{ba} - ig_c \left(E_c^+ e^{i\mathbf{k}_c \cdot \mathbf{r}} + E_c^- e^{-i\mathbf{k}_c \cdot \mathbf{r}}\right) \sigma_{ca} - \Upsilon_a \sigma_{aa} \nonumber \\
		\dot{\sigma}_{bb} &= ig_p E_p e^{i\mathbf{k}_p \cdot \mathbf{r}} \sigma_{ba} + \Upsilon_b \sigma_{aa} \nonumber \\
		\dot{\sigma}_{cc} &= ig_c \left(E_c^+ e^{i\mathbf{k}_c \cdot \mathbf{r}} + E_c^- e^{-i\mathbf{k}_c \cdot \mathbf{r}}\right) \sigma_{ca} + \Upsilon_c \sigma_{aa} \tag{2} \\
		\dot{\sigma}_{ba} &= ig_p E_p e^{i\mathbf{k}_p \cdot \mathbf{r}} (\sigma_{bb} - \sigma_{aa}) + ig_c \left(E_c^+ e^{i\mathbf{k}_c \cdot \mathbf{r}} + E_c^- e^{-i\mathbf{k}_c \cdot \mathbf{r}}\right) \sigma_{bc} - \Gamma_{ba} \sigma_{ba} \nonumber \\
		\dot{\sigma}_{ca} &= ig_c \left(E_c^+ e^{i\mathbf{k}_c \cdot \mathbf{r}} + E_c^- e^{-i\mathbf{k}_c \cdot \mathbf{r}}\right) (\sigma_{cc} - \sigma_{aa}) + ig_p E_p e^{i\mathbf{k}_p \cdot \mathbf{r}} \sigma_{bc}^* - \Gamma_{ca} \sigma_{ca} \nonumber \\
		\dot{\sigma}_{bc} &= ig_c \left(E_c^+ e^{i\mathbf{k}_c \cdot \mathbf{r}} + E_c^- e^{-i\mathbf{k}_c \cdot \mathbf{r}}\right) \sigma_{ba} - ig_p E_p e^{i\mathbf{k}_p \cdot \mathbf{r}} \sigma_{ca}^* - \Gamma_{bc} \sigma_{bc} \nonumber
	\end{align}
	
	Given that the probe pulse is significantly weaker than the standing-wave control laser fields, the equations can be solved using a perturbative approach, as outlined in \cite{fleischhauer2002}. In this method, the atomic density matrix elements are expanded in powers of $\epsilon$, defined as the ratio of the amplitudes of the probe and control pulses. Under the initial condition that the majority of atoms are in the ground state ($\sigma_{bb}^{(0)} \to 1$, $\sigma_{\mu\nu}^{(0)} \to 0$ for $\mu\nu \neq bb$), the atomic density matrix elements are obtained as follows:
	\begin{align}
		\sigma_{ba} &= \frac{1}{i\Omega_c^*} \left( \frac{\partial}{\partial t} + \Gamma_{bc} \right) \sigma_{bc} \tag{3a} \\
		\sigma_{bc} &= \frac{-g_p E_p}{\Omega_c} + \frac{1}{i\Omega_c} \left( \frac{\partial}{\partial t} + \Gamma_{ba} \right) \sigma_{ba} \tag{3b} \\
		\sigma_{bc} &= \frac{-g_p E_p}{\Omega_c} + \frac{1}{i\Omega_c} \left( \frac{\partial}{\partial t} + \Gamma_{ba} \right) \frac{1}{i\Omega_c^*} \left( \frac{\partial}{\partial t} + \Gamma_{bc} \right) \sigma_{bc} \tag{3c}
	\end{align}
	Here, $\Omega_c = g_c (E_c^+ e^{i\mathbf{k}_c \cdot \mathbf{r}} + E_c^- e^{-i\mathbf{k}_c \cdot \mathbf{r}}) = \Omega_{c+} e^{i\mathbf{k}_c \cdot \mathbf{r}} + \Omega_{c-} e^{-i\mathbf{k}_c \cdot \mathbf{r}}$, where $\Omega_{c\pm}$ are the control field components. The equations of motion derived in this framework, such as (3a), (3b), and (3c), evolve very slowly. To account for this, we define the characteristic timescale of laser operation as $T_{\text{ch}}$. The slowly varying atomic operator $\sigma_{bc}$ is expanded in powers of $1/(\Gamma_{ba} T_{\text{ch}})$. Expanding up to zeroth order in equation (3c) and substituting $\sigma_{bc}$ into equation (3a), we derive the following:
	\begin{align}
		\sigma_{bc} &= \frac{-g_p E_p}{\Omega_c} \tag{4a}
	\end{align}
	By inserting the value of $\sigma_{bc}$ in equation (4a), we get:
	\begin{align}
		\sigma_{ba} &= -\frac{1}{i\Omega_c^* \Omega_c} \left( \frac{\partial}{\partial t} + \Gamma_{bc} \right) g_p E_p \tag{4b}
	\end{align}
	Rewriting these equations using the definitions $\Omega_c$, we find the spin excitation of the probe where $\Omega = \sqrt{|\Omega_{c+}|^2 + |\Omega_{c-}|^2}$, $\kappa_\pm = \Omega_{c\pm} / \Omega$ is the total Rabi frequency of the control field, and $\phi$ represents a constant phase.
	\begin{align}
		\sigma_{ba} &= -\frac{1}{i\Omega^2(1 + 2\kappa_+\kappa_- \cos(2k_c z + \phi))} \left( \frac{\partial}{\partial t} + \Gamma_{bc} \right) \frac{g_p E_p}{\Omega} \tag{4c}
	\end{align}
	
	We must also know the wave-equation of the electromagnetic field in an atomic medium. It has been shown that the wave equation can be written as \cite{fleischhauer2002}:
	\begin{align}
		\left( \frac{\partial}{\partial t} + c \frac{\partial}{\partial z} \right) E_p^+ &= i N g_p \sigma_{ba}^+ \tag{5}
	\end{align}
	
	Having derived the atomic spin excitation operator for our atomic ensemble, we are now in a position to review in detail how these spin excitations help quantum states to be stored as quantum information. In this study, we consider a system where all atoms are initially prepared in the ground state $|b,0\rangle$. Under this condition, the interaction couples the system exclusively to the totally symmetric Dicke-like states. Specifically, if the field is prepared in an initial state containing at most one photon, the relevant eigenstates of the bare system are as follows: The total ground state $|b,0\rangle$, which remains unaffected by the interaction. The ground state with a single photon in the field, $|b,1\rangle$. The singly excited atomic states $|a,0\rangle$ and $|c,0\rangle$. These states form the basis of the interaction dynamics, with the system transitioning among them under the influence of the laser fields. The exclusion of other states simplifies the analysis and highlights the key physical processes associated with single-photon and single-atom excitations in the system \cite{fleischhauer2002}.
	\begin{align*}
		|b\rangle &= |b_1, b_2, \dots, b_N\rangle, \\
		|a\rangle &= \frac{1}{\sqrt{N}} \sum_{j=1}^N |b_1, \dots, a_j, \dots, b_N\rangle, \\
		|c\rangle &= \frac{1}{\sqrt{N}} \sum_{j=1}^N |b_1, \dots, c_j, \dots, b_N\rangle
	\end{align*}
	The eigenstates of the Hamiltonian in Eq. (1) can be found as
	\begin{align*}
		|D\rangle &= \cos\theta |b\rangle - \sin\theta |c\rangle \\
		|B_+\rangle &= \sin\theta \sin\phi |b\rangle + \cos\theta \sin\phi |c\rangle + \cos\phi |a\rangle \\
		|B_-\rangle &= \sin\theta \cos\phi |b\rangle + \cos\theta \cos\phi |c\rangle - \sin\phi |a\rangle; \quad \tan\theta = \frac{\sqrt{N} g_p}{\Omega_c}, \quad \tan2\phi = \frac{2\sqrt{N g_p^2 + \Omega_c^2}}{\Delta_p}
	\end{align*}
	$|D\rangle$ can be identified as the required dark state, which we already discussed in the Introduction. For the ensemble, the dark state for a single photonic excitation can be given as: $|D,1\rangle = \cos\theta(t) |b,1\rangle - \sin\theta(t) |c,0\rangle$. Here, $\tan\theta$ is defined as the mixing angle. For $n$ such excitations the expression can be given as \cite{fleischhauer2002}:
	\begin{align*}
		|D,n\rangle &= \sum_{k=0}^n \frac{\sqrt{n!}}{\sqrt{k!(n-k)!}} (-\sin\theta)^k (\cos\theta)^{n-k} |c^k, n-k\rangle
	\end{align*}
	By adiabatic rotation of the mixing angle $\theta$ from $0$ to $\pi/2$, it leads to reversible transfer of the photonic number states to the collective atomic spin excitations stored at the final stable state $|c\rangle = \frac{1}{\sqrt{N}} \sum_{j=1}^N |b_1, \dots, c_j, \dots, b_N\rangle$ \cite{fleischhauer2002}.
	\begin{align}
		|D,n\rangle : |b\rangle |n\rangle \to |c^n\rangle |0\rangle \tag{6}
	\end{align}
	The operation generates a tensor product mix of the photonic states and the collective atomic spin excitations as \cite{fleischhauer2002}
	\begin{align*}
		\sum_{n,m} \rho_{nm} |n\rangle\langle m| \otimes |b\rangle\langle b| \to |0\rangle\langle 0| \otimes \sum_{n,m} \rho_{nm} |c^n\rangle\langle c^m|
	\end{align*}
	
	The formulation presented above is based on the time-independent version of the Hamiltonian in Eq. (1); however, to derive the propagation equations of the probe inside the atomic ensemble with the preserved quantum states in the atomic excitation, we need to work with the time-dependent dynamics of the above system, featuring a time-dependent control field. Consequently, the atomic excitation derived in Eqs. 4(a)-4(c) will also have to be time-dependent. Please note we need to solve the coupled equations (4(c)) and (5) to solve the propagation problem. Physically, this means we calculate how the atomic excitation evolves as the electromagnetic field of the probe propagates inside the atomic ensemble. To perform the time-dependent calculations effectively, we need to arrive at a single equation which can predict the time dependence of both the atomic operator and the electromagnetic field. To solve the problem, a quasi-particle called a `dark state polariton' was introduced in \cite{fleischhauer2002}. The quantum field $\Psi$ is defined as the linear superposition of the bosonic field (in our case probe laser electromagnetic field $E_p$) and an atomic excitation operator field $\sigma_{bc}$ allowing the above dark state transitions.
	\begin{align}
		\Psi_+ &= \cos\theta E_p^+ - \sin\theta \sqrt{N} g_p \sigma_{bc} \tag{7}
	\end{align}
	These field operators follow bosonic commutation relations.
	\begin{align*}
		[\Psi_k, \Psi_{k'}^\dagger] &\approx \delta_{k,k'}
	\end{align*}
	The dark state transition then can be written as $|n_k\rangle = \frac{1}{\sqrt{n!}} (\Psi_k^\dagger)^n |0\rangle |b_1 \dots b_N\rangle$, where $|0\rangle$ is the vacuum state, i.e.: $H_0 = 0$. Now we are ready to derive the propagation equations for the polariton fields. For including the atomic diffusion into the polariton picture, the propagation equation for the dark polariton field needs to be of second order in spatial coordinate. For that we expand 3(c) up to first order in $\Gamma_{ba}/T_{\text{ch}}$, and then insert the value of $\sigma_{bc}$ in 4(a), we thus get
	\begin{align}
		\sigma_{ba}^+ &= -\frac{1}{i\Omega_c^2} \frac{\partial}{\partial t} (g_p E_p) + \frac{\Gamma_{ba}}{\Omega_c^4} \frac{\partial^2}{\partial t^2} (g_p E_p) \tag{8}
	\end{align}
	By using the definition of the polariton field in (7) we get
	\begin{align}
		\sqrt{N} \sigma_{ba}^+ &= -\frac{\sin\theta}{i\Omega(1 + 2\kappa_+\kappa_- \cos(2k_c z + \phi))} \frac{\partial}{\partial t} (\Psi_+ e^{ik_p z}) \nonumber \\
		&\quad + \frac{\Gamma_{ba} \sin\theta \tan^2\theta g_p^2 N}{[i\Omega(1 + 2\kappa_+\kappa_- \cos(2k_c z + \phi))]^2} \frac{\partial^2}{\partial t^2} (\Psi_+ e^{ik_p z}) \tag{9}
	\end{align}
	The wave equation for the probe field then can be written in terms of the dark polariton field, considering a constant mixing angle ($\dot{\theta} \approx 0$) as:
	\begin{align}
		\left( \frac{\partial}{\partial t} + c \cos^2\theta \frac{\partial}{\partial z} \right) \Psi_+ &= -\frac{\sin\theta}{i\Omega(1 + 2\kappa_+\kappa_- \cos(2k_c z + \phi))} \frac{\partial}{\partial t} (\Psi_+ e^{ik_p z}) \nonumber \\
		&\quad + \frac{\Gamma_{ba} \sin\theta \tan^2\theta g_p^2 N}{[i\Omega(1 + 2\kappa_+\kappa_- \cos(2k_c z + \phi))]^2} \frac{\partial^2}{\partial t^2} (\Psi_+ e^{ik_p z}) \tag{10}
	\end{align}
	
	In a solid-state system, the atomic operators exhibit vibrations corresponding to all possible wave vectors, enabling their representation as a Fourier series. These vibrations correspond to the complete set of standing wave modes of the control field within the cavity. The interaction between the atomic spin excitation field operators and these modes gives rise to polaritons, which encode the quantum state of the probe pulse photons. Therefore, the denominator $\frac{1}{i\Omega(1 + 2\kappa_+\kappa_- \cos(2k_c z + \phi))}$ appearing on the right-hand side of equation (10) can similarly be expressed as a Fourier series.
	\begin{align}
		\frac{1}{1 + 2\kappa_+\kappa_- \cos(2k_c z + \phi)} &= \frac{a_0}{2} + \sum_{n=1}^\infty a_n \cos(nx) \tag{11a} \\
		\frac{1}{(1 + 2\kappa_+\kappa_- \cos(2k_c z + \phi))^2} &= \frac{b_0}{2} + \sum_{n=1}^\infty b_n \cos(nx) \tag{11b}
	\end{align}
	Substituting the Fourier expansion of the denominator in (10) we get:
	\begin{align}
		\left( \frac{\partial}{\partial t} + c \cos^2\theta \frac{\partial}{\partial z} \right) \Psi_+ &= -\frac{\sin\theta}{i\Omega} \left[ \frac{a_0}{2} + \sum_{n=1}^\infty a_n \cos(nx) \right] \frac{\partial}{\partial t} (\Psi_+ e^{ik_p z}) \nonumber \\
		&\quad + \frac{\Gamma_{ba} \sin\theta \tan^2\theta g_p^2 N}{(i\Omega)^2} \left[ \frac{b_0}{2} + \sum_{n=1}^\infty b_n \cos(nx) \right] \frac{\partial^2}{\partial t^2} (\Psi_+ e^{ik_p z}) \tag{12}
	\end{align}
	The propagation equation for the spin excitation operator is derived by finding the correct Fourier coefficient, for which the unidirectional polariton component $\Psi_+$ has the corresponding primary Fourier mode of the spin excitation (i.e., $\sigma_{ba}^+$). The Fourier coefficients are given as \cite{hansen2007}:
	\begin{align}
		a_n &= \frac{1}{\pi} \int_0^{2\pi} \frac{\cos(nx)}{1 + y\cos x} dx, \quad y = 2\kappa_+\kappa_- \tag{13a} \\
		b_n &= \frac{1}{\pi} \int_0^{2\pi} \frac{\cos(nx)}{(1 + y\cos x)^2} dx, \quad x = 2k_c z + \phi, \quad y = 2\kappa_+\kappa_- \tag{13b}
	\end{align}
	The final propagation wave equation for the travelling probe obtained after evaluating Eq. (12) further, considering a very low decoherence rate $\Gamma_{bc} \approx 0$ and using the relation between the second order derivative of space and time coordinates $\frac{\partial^2}{\partial t^2}\Psi_+ = c^2(\cos\theta')^4 \frac{\partial^2}{\partial z^2}\Psi_+$, will be as follows:
	\begin{align}
		\left( \frac{\partial}{\partial t} + c(\cos\theta')^2 \frac{\partial}{\partial z} \right) \Psi_+ &= \Gamma_{ba} g_p^2 N \sin^2\theta' \tan^2\theta' \frac{\partial^2}{\partial z^2} \left( \frac{b_n}{a_n} \Psi_+ \right) \quad \text{where } \theta' = \tan^{-1} \left( \frac{a_0}{2} \tan\theta e^{i N_2 \phi} \right) \tag{14}
	\end{align}
	As we have chosen a stationary atomic system, only first order Fourier coefficients are of significance as they are much larger than the higher order Fourier coefficients, so we only need to find the $a_n = a_0, b_n = b_0$ as we are considering the special case of $k_c = k_p = k$ using Eq. (13a) and (13b): $a_0 = \frac{2}{\sqrt{1-y^2}}$, $b_0 = \frac{2}{(1-y^2)^{3/2}}$. Substituting the Fourier coefficients in Eq.(14), approximating the group velocity $\cos^2\theta' \approx \sqrt{1-y^2} \cos^2\theta$ and evaluating we get:
	\begin{align}
		\left( \frac{\partial}{\partial t} + c \sqrt{1-y^2} \cos^2\theta \frac{\partial}{\partial z} \right) \Psi_+ &= \frac{\Gamma_{ba} g_p^2 N c^2 \cos^4\theta}{1-y^2} \frac{\partial^2}{\partial z^2} \Psi_+ \tag{15}
	\end{align}
	
	The wave equation governing the polariton fields under non-adiabatic conditions is presented, incorporating a second-order derivative term on the right-hand side. This term, identified as the diffusion term, arises from pulse broadening, which leads to the loss of polariton fields stored within the material as quantum memory. The term diminishes due to the standing wave control field Rabi frequencies. The equation is solved using the Fourier transform method, yielding a solution in Fourier space after integrating both sides. Assuming an initial Gaussian probe wave, expressed as $\Psi_+(w,0) = \Psi_{+0} e^{-w^2/2}$, selecting units where $\Gamma_{ba} g_p^2 N = c = \cos^2\theta_0 = 1$ for simplicity and choosing the group velocity's time dependence as $v_g = c(\cos\theta)^2 = c \cos^2\theta_0 \tanh(t/T_s)$, where $T_s$ is the characteristic switching time of the laser, the solution is subsequently transformed back to real space via an inverse Fourier transform. The resulting expression, obtained numerically, provides insight into the behavior of the polariton field under the given conditions:
	\begin{align}
		\Psi_+(z,t) &= \Psi_{+0} \exp\left[ -\frac{\left(z - \sqrt{1-y^2} \ln(\cosh t)\right)^2}{2\left( 1 + \frac{2(t - \tanh t)}{(1-y^2) T_s} \right)} \right] \left( 1 + \frac{2(t - \tanh t)}{(1-y^2) T_s} \right)^{-1/4} \tag{16}
	\end{align}
	The total energy density obtained after substituting $\Psi_+(z,t)$ in terms of field amplitude $E_p$ based on the definition (7) is:
	\begin{align}
		U_p(z,t) &= \frac{\hbar\omega_p}{V} |E_p|^2 = \frac{\hbar\omega_p}{V} \cos^2\theta_0 \tanh\left(\frac{t}{T_s}\right) |\Psi_{+0}|^2 \nonumber \\
		&\quad \times \exp\left[ -\frac{\left(z - \sqrt{1-y^2} \ln(\cosh t)\right)^2}{1 + \frac{2(t - \tanh t)}{(1-y^2) T_s}} \right] \left( 1 + \frac{2(t - \tanh t)}{(1-y^2) T_s} \right)^{-1/2} \tag{17}
	\end{align}
	The total energy density derived under the given conditions reveals that, for a perfect standing wave configuration $\kappa_+ = \kappa_- = 1/\sqrt{2}$ or $y=1$, the diffusion term on the right-hand side of the wave equation vanishes entirely.
	\begin{align}
		U_p(z,t) &= \frac{\hbar\omega_p}{V} \cos^2\theta_0 \tanh\left(\frac{t}{T_s}\right) |\Psi_{+0}|^2 \exp(-z^2/2) \tag{18}
	\end{align}
	In this scenario, there is no energy loss attributable to diffusion, although losses arising from decoherence may still persist. This result highlights the significance of the standing wave configuration in minimizing diffusion-related losses in the system. This feature has been previously demonstrated for a standing wave probe pulse trapped inside an EIT cavity. Our results extend this understanding by showing that diffusion losses of a traveling probe pulse inside an EIT cavity can also be minimized by the primary modes of the standing wave control field. The application of the standing wave control field reduces the traveling probe field losses, leading to extended coherence and storage times. However, our formulation assumes that the spontaneous downward transition of atoms in the dark state is negligible compared to the atomic population at that level. In a real atomic system, the spontaneous downward transition is significant and must be considered. This transition follows an exponential decay law, and incorporating the loss due to spontaneous emission allows the energy density derived earlier to be approximately recalculated as:
	\begin{align}
		U_p(z,t) &\approx \frac{\hbar\omega_p}{V} \cos^2\theta_0 \tanh\left(\frac{t}{T_s}\right) |\Psi_{+0}|^2 \nonumber \\
		&\quad \times \exp\left[ -\frac{\left(z - \sqrt{1-y^2} \ln(\cosh t)\right)^2}{1 + \frac{2(t - \tanh t)}{(1-y^2) T_s}} \right] \left( 1 + \frac{2(t - \tanh t)}{(1-y^2) T_s} \right)^{-1/2} \exp(-\Gamma_{bc} t) \tag{19}
	\end{align}

	\section{Proposed Experimental Validation}
	
	Several experiments have demonstrated the use of standing wave control fields to create an Electromagnetically Induced Transparency (EIT) window for standing wave pulses in atomic systems. However, only a few studies have focused on stationary solid-state atomic systems. Our proposed experiment aims to explore EIT-based quantum memory storage in a cryogenically cooled (4.5 K) rare-earth-doped $\text{Pr}^{3+}:\text{Y}_2\text{SiO}_5$ crystal. Building on the work of G. Heinze et al. (2013) and D. Schraft et al. (2016), who achieved high storage efficiencies (optical depth $\sim 96$) and extended storage times (up to one minute) using traveling wave control fields, we propose a new approach utilizing a standing wave control field. Specifically, we will employ two counter-propagating control laser fields through the crystal to create a standing wave pattern \cite{heinze2013, schraft2016}. 
	\begin{figure}[htbp]
		\centering
		\includegraphics[width=0.7\linewidth]{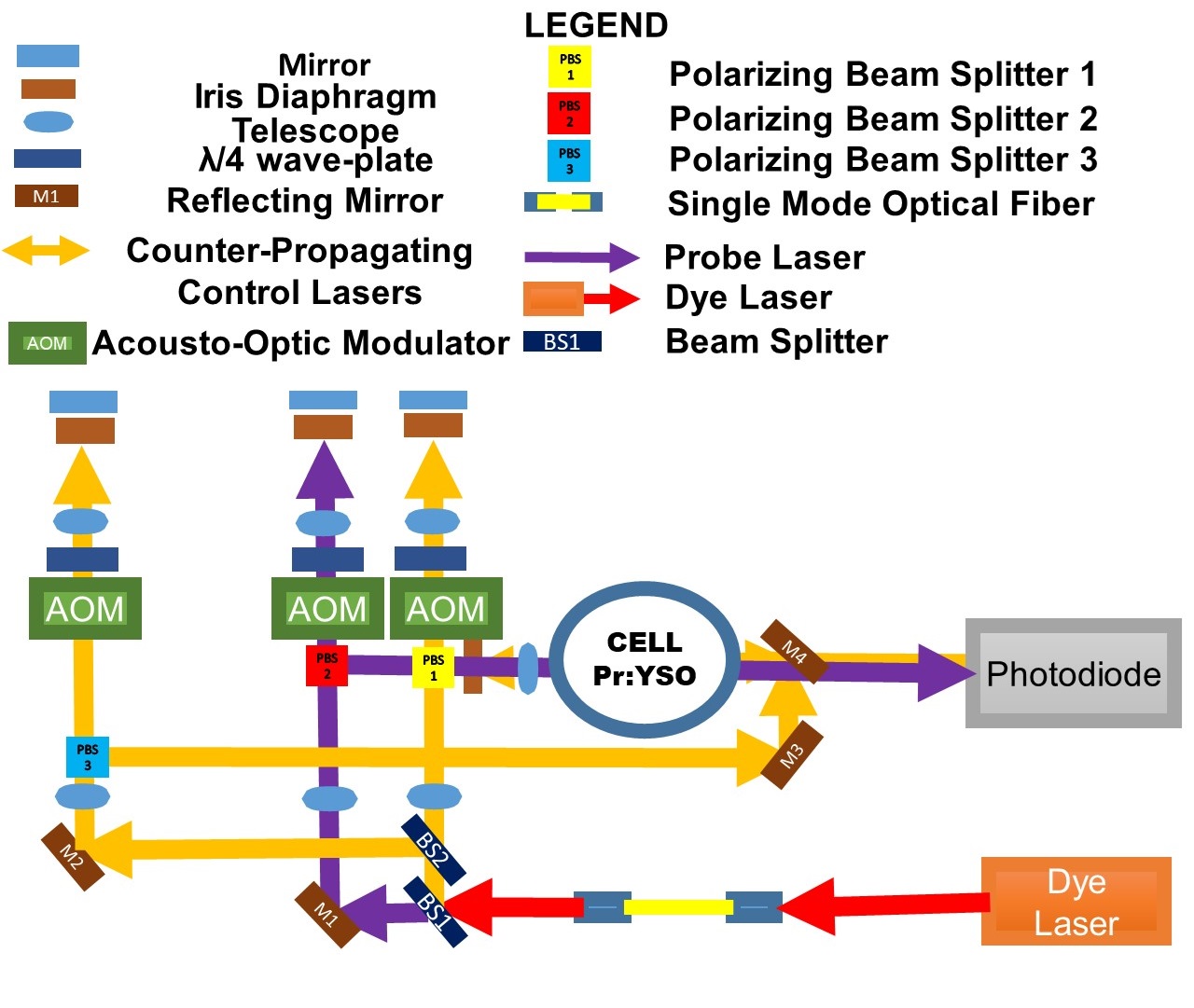}
		\caption{Proposed experimental setup for standing wave control field EIT-based quantum memory storage.}
		\label{fig:fig2}
	\end{figure}
	The experiment uses the hyperfine levels $^3\text{H}_4 \leftrightarrow ^1\text{D}_2$ of $\text{Pr}^{3+}:\text{Y}_2\text{SiO}_5$ to form a $\Lambda$-level EIT system. A frequency-stabilized laser with a bandwidth below 100 kHz (FWHM) on a 100 ms timescale will generate preparation, control, and probe pulses. Acousto-optic modulators (AOMs) will modulate pulse intensities and frequencies. The preparation sequence involves optical pumping and repumping to prepare the atomic population in the ground state. A preparation pulse with a spectral range of $0 < \Delta\nu < 18 \text{ MHz}$ and a duration of 84 ms will create a spectral pit in the absorption medium. Repumping the two hyperfine components of the ground state $^3\text{H}_4$ will involve a short pulse of 30 ms duration with a relative frequency of +25.7 MHz. A cleaning pulse of 100 $\mu$s will remove residual excited-state population \cite{beil2008, nilsson2004}. 
	
	The experimental setup, as shown in Fig.~\ref{fig:fig2}, uses probe and control pulses derived from a dye laser, passed through single-mode optical fibers. A beam splitter (BS1) separates the pulses into a weak probe beam and a control beam. The control beam is further split using another beam splitter (BS2) to generate a standing wave control field. The probe pulse is directed onto the $\text{Pr}^{3+}:\text{Y}_2\text{SiO}_5$ crystal using reflective mirrors and alignment telescopes, while the counter-propagating control beams focus on the crystal with minimal angular separation. Polarizing beam splitters (PBS) and AOMs polarize and fine-tune the beams' intensity and frequency. Initially, only the control beams are applied to create a standing wave within the crystal. After a microsecond delay, the probe pulse is introduced, inducing EIT in the medium. This reduces the group velocity of the probe to near zero, forming a dark-state polariton—a coherent superposition of atomic states. The standing wave control field minimizes diffusion losses, enabling efficient quantum memory storage within the adiabatic time limit. The storage duration is governed by the hyperfine level lifetimes, with the stored probe pulse retrievable by turning off the control field, completing the memory cycle.
	
	\section{Results}
	
	Equation (16) highlights the critical role of the Rabi frequencies of counter-propagating control lasers in shaping the traveling dark-state polariton (DSP) waveform. The counter-propagating laser pulses generate cavity modes that are inherently dependent on these Rabi frequencies, which, in turn, influence the EIT window through which the traveling probe polariton propagates. By varying the control laser fields, the Fourier modes of the control lasers can be adjusted, allowing precise manipulation of the probe pulse’s shape and duration. This tunability facilitates the manipulation of photonic quantum states within the atomic medium as polaritons, making them highly suitable for quantum memory applications that require extended coherence times. Different ratios of $\Omega_{c\pm}$ result in either enhanced or diminished waveforms of the probe polariton field. Notably, when counter-propagating control fields of equal intensities are applied, the polariton field stored within the crystal experiences no diffusion loss. Under these optimal conditions, the adiabatic storage of the probe pulse is limited solely by the spontaneous emission rates of the ground-level hyperfine states $|c\rangle$ and $|b\rangle$, ensuring high-fidelity quantum memory storage. 
	
	\begin{figure*}[htbp]
		\centering
		\includegraphics[width=0.9\textwidth]{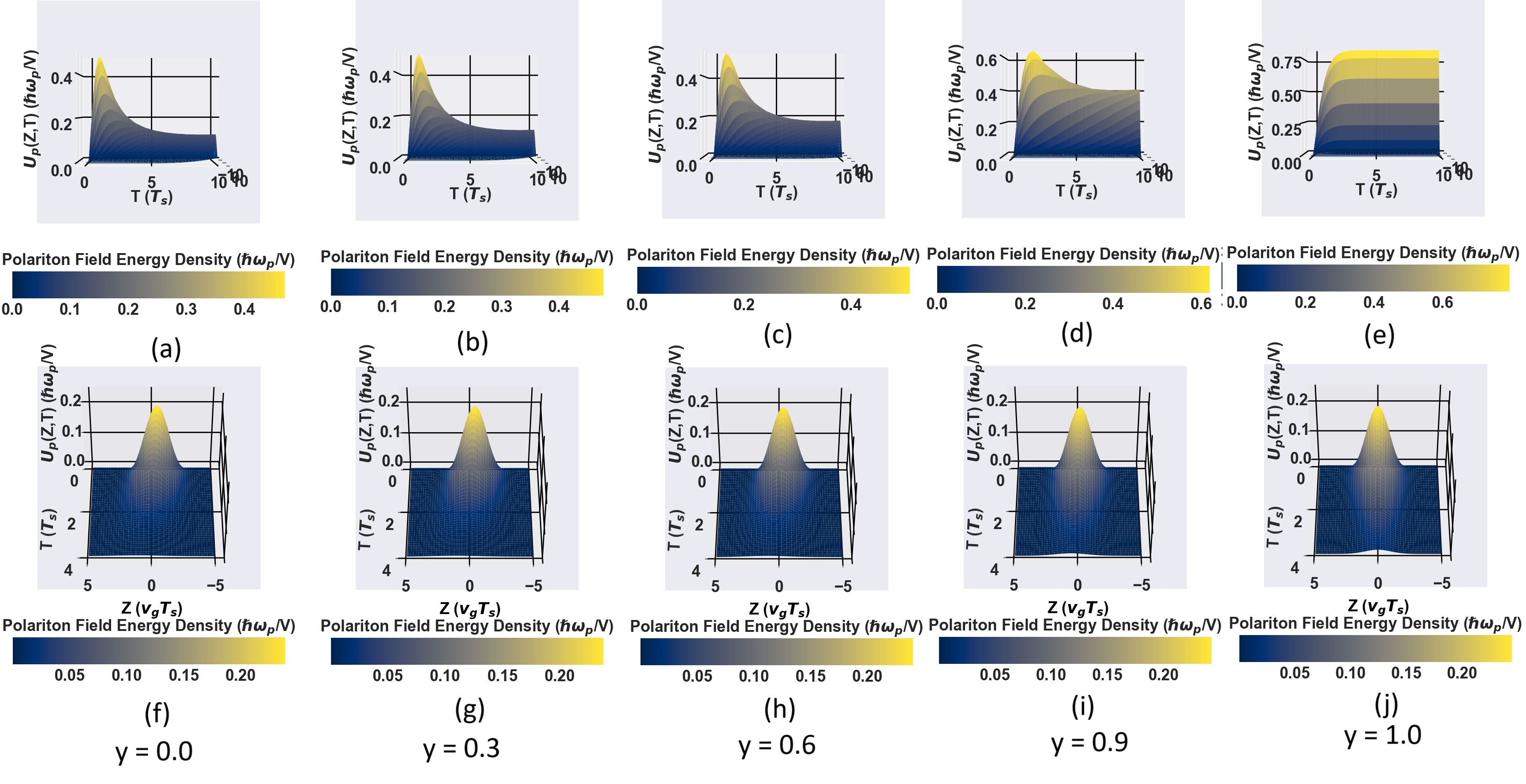}
		\caption{(a)-(e) Plots of total polariton field energy density $U_p(z,t)$ of the probe pulse showing the temporal variation in arbitrary units without considering spontaneous emission loss ($\Gamma_{bc} \approx 0$). (f)-(j) Plots of total polariton field energy density $U_p(z,t)$ of the probe pulse showing the pulse broadening in spatial units, considering spontaneous emission loss. The plots range from $y = 0.0$ (travelling wave control field case) up to $y = 1.0$ (perfectly standing wave control field case). Here, $y = 2\kappa_+\kappa_-$. We have chosen the constants: $\Gamma_{ba} N g_p^2 = \cos^2\theta_0 = c = 1$, time $T$ is in units of $T_s$, distance $z$ is in units of $v_g T_s$, and energy is in units of $\hbar\omega_p/V$.}
		\label{fig:fig3}
	\end{figure*}
	
	Fig.~\ref{fig:fig3} (a)-(e) illustrates the energy decay of the polariton field under various control field configurations. In the case of a traveling wave control field ($y=0.0$), the polariton energy exhibits exponential decay due to diffusion losses. With counter-propagating control fields ($y=0.3-0.9$), the diffusion losses are significantly reduced, resulting in a slower decay of the polariton field energy and an extended coherence time for the photonic states mapped into atomic states. For a perfectly standing wave control field ($y=1.0$), diffusion losses are completely eliminated, and the polariton energy remains constant over time in the absence of spontaneous emission. This configuration ensures that the collective atomic excitations remain undisturbed, theoretically enabling indefinite storage. However, in real atomic systems, the finite spontaneous emission rate imposes a practical limitation on the storage duration. 
	
	Fig.~\ref{fig:fig3} (f)-(j) illustrates the effect of the standing wave control on the pulse broadening of the probe pulse inside the medium. As we approach the perfectly standing probe pulse condition, pulse broadening decreases and gets eliminated completely when a perfectly standing wave control field is obtained.
	
	\begin{figure}[htbp]
		\centering
		\includegraphics[width=0.7\linewidth]{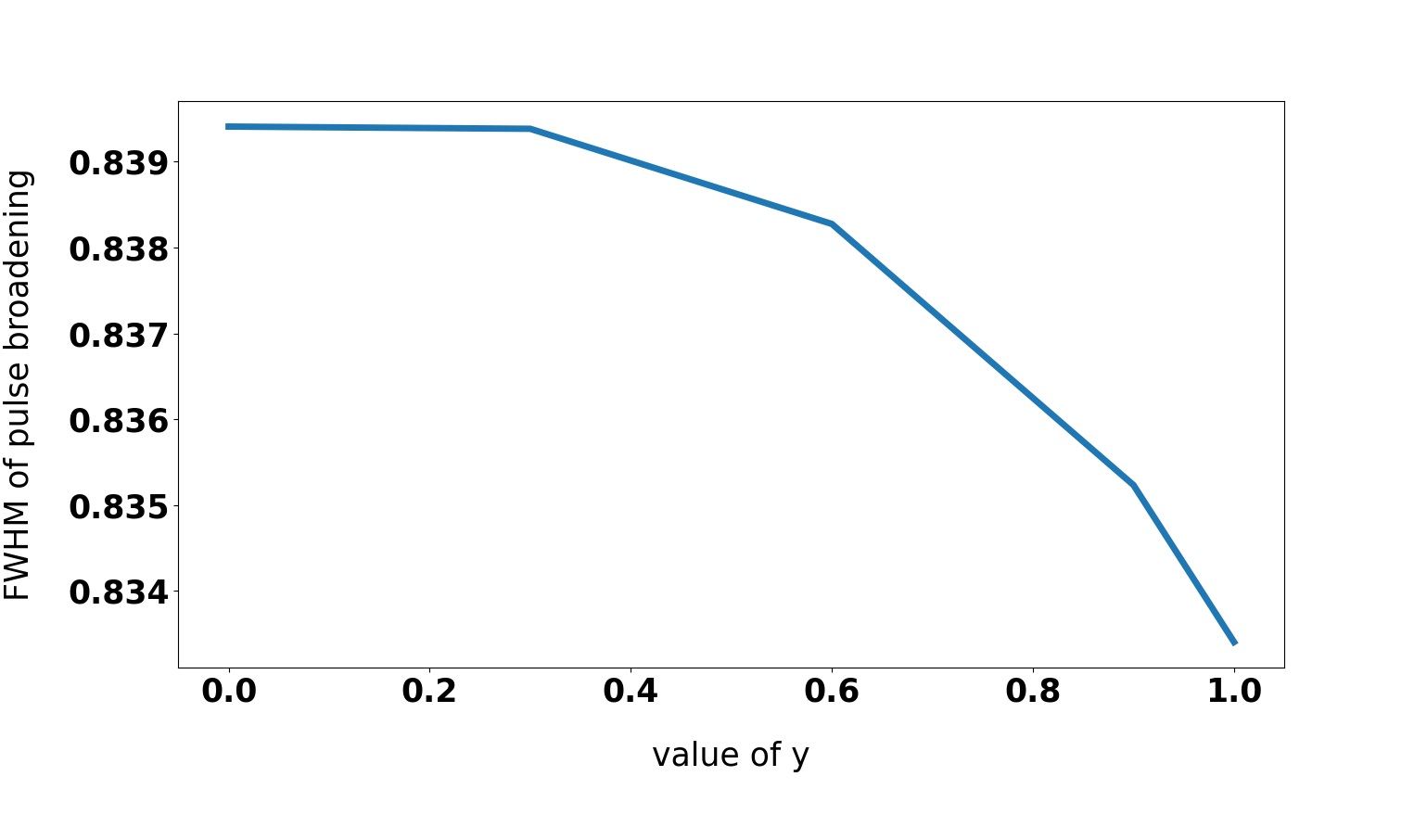}
		\caption{Plots of FWHM of the pulse broadening against product of the Rabi frequencies of the counter-propagating fields, $y=2\kappa_+\kappa_-$, showing as we approach perfectly standing wave case, we minimise the pulse broadening, due to the elimination of the atomic diffusion by the cavity modes, generated by the standing wave control laser field.}
		\label{fig:fig4}
	\end{figure}
	
	Figure \ref{fig:fig4} illustrates the variation in pulse broadening of the polariton with the Rabi frequency ratio, $y=2\kappa_+\kappa_-$. In the traveling wave control field configuration ($y=0$), the Full Width at Half Maximum (FWHM) of the polariton energy density distribution is the largest, indicating significant pulse broadening caused by diffusion-induced expansion. In contrast, for counter-propagating control fields ($y=0.3-0.9$), the FWHM decreases as diffusion losses are reduced, resulting in a narrower pulse width. The perfectly standing wave control field case ($y=1$) demonstrates the smallest FWHM, reflecting minimal pulse broadening due to the complete elimination of diffusion loss.
	
	\begin{figure}[htbp]
		\centering
		\includegraphics[width=0.7\linewidth]{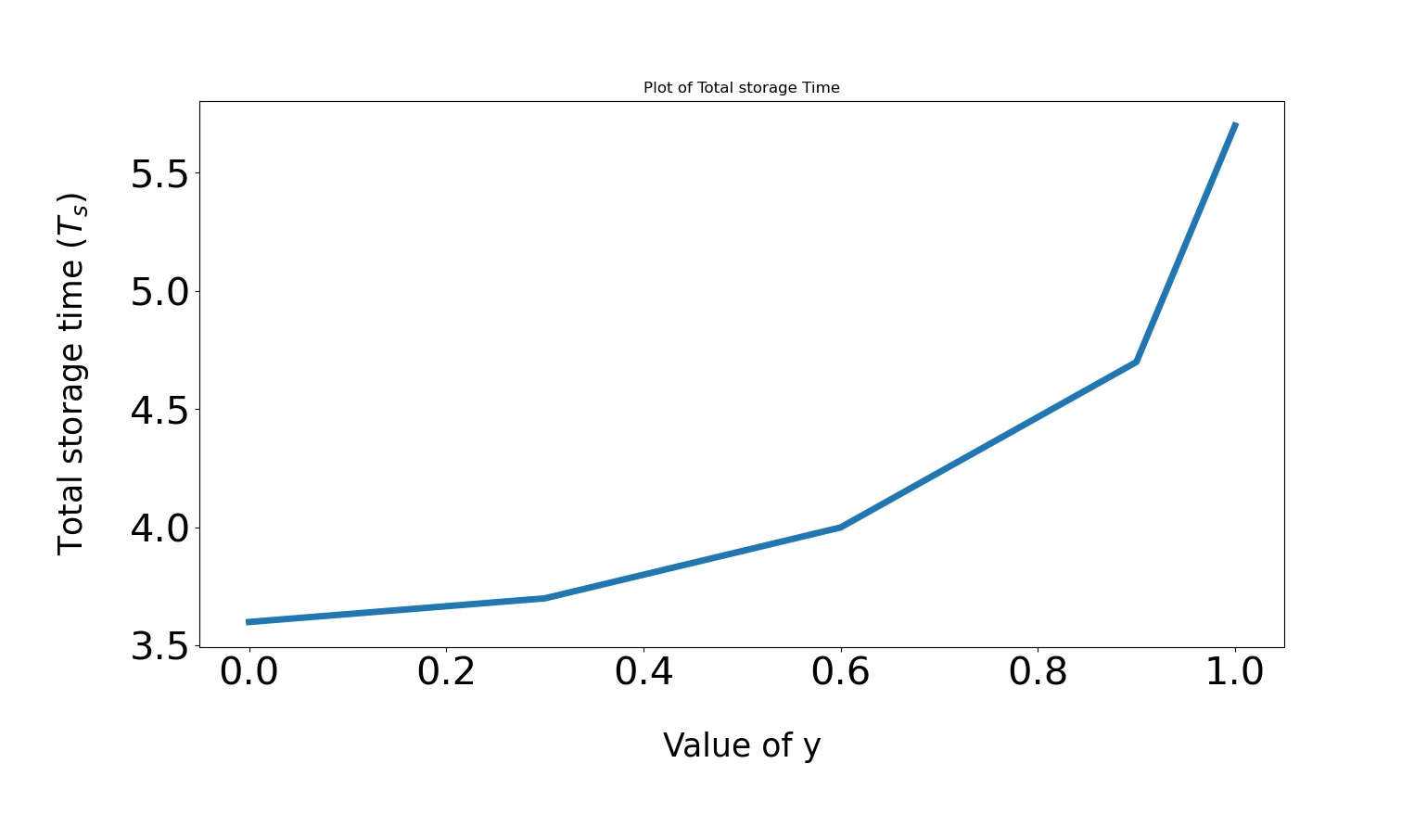}
		\caption{Plot showing the total storage time of the stored polariton field $U_p(0,t)$ of the probe pulse for different values of $y=2\kappa_+\kappa_-$, in arbitrary units, based on the time taken for the energy density $U_p(0,t)$ to fall below $0.01 \hbar\omega_p/V$. We have chosen the constants: $\Gamma_{ba} N g_p^2 = \cos^2\theta_0 = c = 1$, time $t$ is in units of $T_s$, and energy is in units of $\hbar\omega_p/V$. Note how the standing wave case ($y = 1$) furnishes us with a prolonged storage time, only limited by the lifetime of the energy levels.}
		\label{fig:fig5}
	\end{figure}
	
	Figure \ref{fig:fig5} illustrates the perfectly standing wave control field ($\kappa_+ = \kappa_- = 1/\sqrt{2}$ or $y=1$); the storage duration of the probe pulse is significantly extended compared to the traveling wave case ($\kappa_+ = 1, \kappa_- = 0$ or $y=0$), based on the time taken for the energy density $U_p(0,t)$ to fall below $0.01 \hbar\omega_p/V$. The storage time increases from approximately $T \sim 3.6 T_s$ for the traveling wave case to $T \sim 5.7 T_s$ for the standing wave case. Here, $T_s$ represents the switching time of the control laser. These results highlight the advantages of the standing wave control field, which significantly prolongs the storage time of the probe field within the crystal, limited only by the lifetime of the ground state $^3\text{H}_4$. This makes the standing wave configuration a superior choice for quantum memory applications.
	
	\section{Conclusion}
	
	Quantum computers are being envisioned—and increasingly applied—in a wide array of fields, from the use of quantum cryptography in cybersecurity to the potential for quantum simulations in quantum chemistry for chemical engineering applications \cite{bova2021}. One of the most intriguing prospects that has recently captured researchers' attention is the scalability of machine learning models to unprecedented levels, enabled by the significantly faster learning capabilities of quantum computers compared to classical systems \cite{aaronson2015, harrow2009, lloyd2013}. However, the realization of such advanced quantum algorithms relies critically on the availability of reliable, robust, and long-term quantum state storage solutions, which are essential for quantum computing to make a transformative impact on the computational landscape. 
	
	The quantum memory storage process by utilization of the transparency window provided by electromagnetically induced transparency, as described in this paper, can prove to be very effective for such purpose. One of the other potential areas of research in the future is processing of exponential amount of classical information by quantum computers. A potent solution to such a problem has already been provided by Giovannetti, in which he proposed a quantum version of volatile short-term memory such as classical RAM can be realized using various procedures involving phase gates, qutrits, and entanglement \cite{giovannetti2008_1, giovannetti2008_2, giovannetti2008_3}. Quantum memory using trapped photonic states inside an EIT medium can prove very effective in realizing memory cells in such a Q-RAM scheme, pertaining to the quality of photonic states being much less susceptible to environmental interference and being easy to manipulate. Also, quantum state retrieval using a probe pulse from the address shells in Q-RAM schemes, built upon multiple EIT systems, can feature greater fidelity. 
	
	Our analysis demonstrates that standing wave control laser fields significantly enhance the coherence times of traveling probe laser pulses in solid crystalline atomic systems, enabling extended storage durations for the quantum state of photons. Notably, these storage times are independent of the intrinsic properties of the solid-state atomic medium and are instead governed by controllable parameters, such as the ratio of the Rabi frequencies of the counter-propagating control lasers. While prior studies primarily focused on eliminating diffusion losses and improving coherence times for stationary pulses, our work extends these benefits to traveling probe pulses. The observed increase in coherence time, as evidenced by the temporal profile $U_p(z,t)$, results from the broader absorptive range of the control field wave vector in the standing wave configuration compared to the traveling wave case. 
	
	We have also proposed an experimental setup to implement the standing wave control laser field EIT system and investigate the storage of traveling probe pulses in a cryogenically cooled (4.5 K) $\text{Pr}:\text{YSO}$ crystal. While the storage of stationary pulses has been extensively studied in warm and cold atomic gases, the application of counter-propagating control fields to store traveling pulses in solid-state media remains largely unexplored. Recently, Hain et al. (2022) demonstrated the storage of a traveling photon pulse using a traveling wave control field for up to 100 seconds in the $^3\text{H}_4 \leftrightarrow ^1\text{D}_2$ levels of $\text{Pr}:\text{YSO}$ \cite{hain2022}. Building upon this work, our proposed experiment predicts longer coherence times for stored pulses using standing wave control fields, as supported by our theoretical model. Despite these advancements, storage durations remain fundamentally constrained by spontaneous emission. Future research must explore innovative strategies to mitigate these limitations and extend the storage capabilities of optical memories beyond current theoretical boundaries. Addressing these challenges could significantly enhance the performance of optical memories, making them practical for a wider range of applications in quantum technologies.
	
	\begin{acknowledgments}
		Authors would like to thank their host institutions and their respective funding agencies.
	\end{acknowledgments}
	
	\vspace{0.8em}
	\noindent\textbf{Data Availability:} Not applicable.
	
	\vspace{0.5em}
	\noindent\textbf{Conflict of Interest:} The authors have no relevant financial or non-financial interests to disclose.
	
	\vspace{0.5em}
	\noindent\textbf{Authors' Contribution:} All authors contributed equally.
	
	\bibliographystyle{apsrev4-2}
	\bibliography{Travelling_Dark_State_Polariton_as_a_Viable_Quantum_Memory_in_a_Solid-State_Medium_NCNMO}

@article{krylov2021,
	title={Quantum computing and quantum information storage},
	author={Krylov, A. I. and Doyle, J. and Ni, K.-K.},
	journal={Phys. Chem. Chem. Phys.},
	volume={23},
	number={11},
	pages={6341--6343},
	year={2021}
}

@article{feynman1982,
	title={Simulating physics with computers},
	author={Feynman, R. P.},
	journal={International Journal of Theoretical Physics},
	volume={21},
	pages={467--488},
	year={1982}
}

@article{obrien2003,
	title={Demonstration of an all-optical quantum controlled-NOT gate},
	author={O'Brien, J. L. and Pryde, G. J. and White, A. G. and Ralph, T. C. and Branning, D.},
	journal={Nature},
	volume={426},
	pages={264--267},
	year={2003}
}

@article{petrosyan2005,
	title={Towards deterministic optical quantum computation with coherently driven atomic ensembles},
	author={Petrosyan, D.},
	journal={Journal of Optics B: Quantum and Semiclassical Optics},
	volume={7},
	pages={S141},
	year={2005}
}

@article{parkins1993,
	title={Synthesis of arbitrary quantum states via adiabatic transfer of Zeeman coherence},
	author={Parkins, A. S. and Marte, P. and Zoller, P. and Kimble, H. J.},
	journal={Phys. Rev. Lett.},
	volume={71},
	number={19},
	pages={3095--3098},
	year={1993}
}

@article{bergmann1998,
	title={Coherent population transfer among quantum states of atoms and molecules},
	author={Bergmann, K. and Theuer, H. and Shore, B. W.},
	journal={Rev. Mod. Phys.},
	volume={70},
	number={3},
	pages={1003--1025},
	year={1998}
}

@article{fleischhauer2000,
	title={Dark-State Polaritons in Electromagnetically Induced Transparency},
	author={Fleischhauer, M. and Lukin, M. D.},
	journal={Phys. Rev. Lett.},
	volume={84},
	number={22},
	pages={5094--5097},
	year={2000}
}

@article{andre2002,
	title={Manipulating Light Pulses via Dynamically Controlled Photonic Band gap},
	author={Andr{\'e}, A. and Lukin, M. D.},
	journal={Phys. Rev. Lett.},
	volume={89},
	number={14},
	pages={143602},
	year={2002}
}

@article{bajcsy2003,
	title={Stationary pulses of light in an atomic medium},
	author={Bajcsy, M. and Zibrov, A. S. and Lukin, M. D.},
	journal={Nature},
	volume={426},
	pages={638--641},
	year={2003}
}

@article{zimmer2006,
	title={Coherent control of stationary light pulses},
	author={Zimmer, F. E. and Andr{\'e}, A. and Lukin, M. D. and Fleischhauer, M.},
	journal={Optics Communications},
	volume={264},
	pages={441--453},
	year={2006}
}

@article{andre2005,
	title={Nonlinear Optics with Stationary Pulses of Light},
	author={Andr{\'e}, A. and Bajcsy, M. and Zibrov, A. S. and Lukin, M. D.},
	journal={Phys. Rev. Lett.},
	volume={94},
	number={6},
	pages={063902},
	year={2005}
}

@article{hansen2007,
	title={Trapping of light pulses in ensembles of stationary Lambda atoms},
	author={Hansen, K. R. and M{\o}lmer, K.},
	journal={Phys. Rev. A},
	volume={75},
	number={5},
	pages={053802},
	year={2007}
}

@article{fleischhauer2002,
	title={Quantum memory for photons: Dark-state polaritons},
	author={Fleischhauer, M. and Lukin, M. D.},
	journal={Phys. Rev. A},
	volume={65},
	number={2},
	pages={022314},
	year={2002}
}

@article{heinze2013,
	title={Stopped Light and Image Storage by Electromagnetically Induced Transparency up to the Regime of One Minute},
	author={Heinze, G. and Hubrich, C. and Halfmann, T.},
	journal={Phys. Rev. Lett.},
	volume={111},
	number={3},
	pages={033601},
	year={2013}
}

@article{schraft2016,
	title={Stopped Light at High Storage Efficiency in a $\text{Pr}^{3+}:\text{Y}_2\text{SiO}_5$ Crystal},
	author={Schraft, D. and Hain, M. and Lorenz, N. and Halfmann, T.},
	journal={Phys. Rev. Lett.},
	volume={116},
	number={7},
	pages={073602},
	year={2016}
}

@article{beil2008,
	title={Electromagnetically induced transparency and retrieval of light pulses in a $\Lambda$-type and a V-type level scheme in $\text{Pr}^{3+}:\text{Y}_2\text{SiO}_5$},
	author={Beil, F. and Klein, J. and Nikoghosyan, G. and Halfmann, T.},
	journal={Journal of Physics B: Atomic, Molecular and Optical Physics},
	volume={41},
	pages={074001},
	year={2008}
}

@article{nilsson2004,
	title={Hole-burning techniques for isolation and study of individual hyperfine transitions in inhomogeneously broadened solids demonstrated in $\text{Pr}^{3+}:\text{Y}_2\text{SiO}_5$},
	author={Nilsson, M. and Rippe, L. and Kr{\"o}ll, S. and Klieber, R. and Suter, D.},
	journal={Phys. Rev. B},
	volume={70},
	number={21},
	pages={214116},
	year={2004}
}

@article{bova2021,
	title={Commercial applications of quantum computing},
	author={Bova, F. and Goldfarb, A. and Melko, R. G.},
	journal={EPJ Quantum Technology},
	volume={8},
	pages={2},
	year={2021}
}

@article{aaronson2015,
	title={Read the fine print},
	author={Aaronson, S.},
	journal={Nature Physics},
	volume={11},
	pages={291--293},
	year={2015}
}

@article{harrow2009,
	title={Quantum Algorithm for Linear Systems of Equations},
	author={Harrow, A. W. and Hassidim, A. and Lloyd, S.},
	journal={Physical Review Letters},
	volume={103},
	pages={150502},
	year={2009}
}

@misc{lloyd2013,
	title={Quantum algorithms for supervised and unsupervised machine learning},
	author={Lloyd, S. and Mohseni, M. and Rebentrost, P.},
	howpublished={arXiv: Quantum Physics},
	year={2013}
}

@article{giovannetti2008_1,
	title={Architectures for a quantum random access memory},
	author={Giovannetti, V. and Lloyd, S. and Maccone, L.},
	journal={Physical Review A},
	volume={78},
	pages={052310},
	year={2008}
}

@article{giovannetti2008_2,
	title={Quantum Private Queries},
	author={Giovannetti, V. and Lloyd, S. and Maccone, L.},
	journal={Physical Review Letters},
	volume={100},
	pages={230502},
	year={2008}
}

@article{giovannetti2008_3,
	title={Quantum Random Access Memory},
	author={Giovannetti, V. and Lloyd, S. and Maccone, L.},
	journal={Physical Review Letters},
	volume={100},
	pages={160501},
	year={2008}
}

@article{hain2022,
	title={Few-photon storage on a second timescale by electromagnetically induced transparency in a doped solid},
	author={Hain, M. and Stabel, M. and Halfmann, T.},
	journal={New Journal of Physics},
	volume={24},
	pages={023012},
	year={2022}
}
	
\end{document}